\documentclass[12pt]{article}

\usepackage{epsf}
\usepackage{latexsym,euscript}
\usepackage[dvips]{graphicx}
\usepackage{amsmath}
\usepackage{amssymb}
\usepackage{cite}

\usepackage{color,soul}

\begin{document}

\begin{center}
{\Large \textbf{$SU(N)$ polynomial integrals and some applications}}

\vspace*{0.6cm}
\textbf{O.~Borisenko\footnote{email: oleg@bitp.kiev.ua}, 
S.~Voloshyn\footnote{email: billy.sunburn@gmail.com}}

\vspace*{0.3cm}
{\large \textit{ Bogolyubov Institute for Theoretical
Physics, National Academy of Sciences of Ukraine, 03143 Kiev, Ukraine}}
\vspace*{0.3cm}

\textbf{V.~Chelnokov\footnote{on leave from Bogolyubov Institute for Theoretical Physics, Kiev; \\ 
\hspace*{0.5cm} email: Volodymyr.Chelnokov@lnf.infn.it}}

\vspace*{0.3cm}
{\large \textit{Istituto Nazionale di Fisica Nucleare,
Gruppo collegato di Cosenza,
I-87036 Arcavacata di Rende, Cosenza, Italy}}

\end{center}

\begin{abstract}
We use the method of the Weingarten functions to evaluate $SU(N)$ 
integrals of the polynomial type. As an application we calculate various one-link 
integrals for lattice gauge and spin $SU(N)$ theories.  
\end{abstract}

\section{Introduction}

Analytical methods of integral calculations in the lattice gauge theories (LGTs) 
had been an important ingredient since the early days of LGT. Many integrals 
appearing in the strong coupling expansion of the LGT can be found in \cite{creutz, lat_rev}. 
Important progress relevant for this paper was made in Refs. \cite{weingarten, bars_green, bars, samuel_un, 
eriksson, brower, balantekin} for $U(N)$ integrals.  
Many $U(N)$ integrals considered here have been studied in Refs. \cite{kontsevich_mod, moerbeke1,  
moerbeke2, orlov1, orlov2, orlov3}.   
Some integrals for $SU(N)$ group can be found in \cite{creutz, wipf, carlsson}.
Last decade has seen a thorough development of the theory 
of the integrals over $U(N)$ group. These integrals are of polynomial type
\begin{equation}
{\cal I}_N(r,s) \ = \  \int_{U(N)}  dU \ \prod_{k=1}^{r} U_{i_k j_k} \prod_{n=1}^{s} U_{m_n l_n}^* 
\label{integral_def}
\end{equation}
and can be expressed through summation over permutations of the group indices.  
The weight of the permutation is given by the Weingarten function whose theory 
has been developed in a series of papers \cite{collins1, collins2, novak, zinn-justin, 
novaes1, novaes2, collins3, zuber}. 
Precisely, the integral is given by 
\begin{equation} 
{\cal I}_N(r,s) \ = \ \delta_{r,s} \ {\cal I}_N(r)\ , 
\label{unintg_def}
\end{equation}
\begin{equation} 
{\cal I}_N(r) \ = \ \sum_{\tau,\sigma\in S_r}  Wg^N(\tau^{-1}\sigma) \prod_{k=1}^{r}
\delta_{i_k,m_{\sigma(k)}} \delta_{j_k,l_{\tau(k)}} \ .
\label{basic_intg}
\end{equation}
$S_r$ is a group of permutations of $r$ elements and  
$Wg^N(\sigma)$ is the Weingarten function which depends only on the length of the cycles 
of a permutation $\sigma$. Its explicit form is given by 
\begin{equation} 
Wg^N(\sigma) = \frac{1}{(r!)^2} \ \sum_{\lambda\vdash r} \ \frac{d^2(\lambda)}{s_{\lambda}(1^N)} \ \chi_{\lambda}(\sigma) \ , 
\label{Wgn_def}
\end{equation}
where $d(\lambda), \chi_{\lambda}(\sigma)$ are the dimension and the character of 
the irreducible representation $\lambda$ of $S_r$. The irreducible representations $\lambda$ are enumerated 
by partitions $\lambda=(\lambda_1,\lambda_2,\cdots,\lambda_{l(\lambda)})$ of $r$, 
{\it i.e.} $\sum_{i=1}^{l(\lambda)} \lambda_i \equiv |\lambda|= r$, 
where $l(\lambda)$ is the length of the partition and $\lambda_1\geq\lambda_2\geq \cdots \geq \lambda_{l(\lambda)}> 0$. 
The sum in (\ref{Wgn_def}) is taken over all $\lambda$ such that $l(\lambda)\leq N$. 
$s_{\lambda}(X)$, $X = (x_1,x_2,\cdots,x_N)$, is the Schur function ($s_{\lambda}(1^N)$ is the dimension 
of $U(N)$ representation). 
To the best of our knowledge, the character expansion of the form (\ref{Wgn_def}), valid for $r\leq N$, had been derived 
for the first time in \cite{samuel_un}, where the recurrence relations for $Wg^N(\sigma)$ were also presented. 
This result was rederived in \cite{collins1}, again for the case $r\leq N$. In Ref.\cite{collins2} it was shown that 
the result (\ref{basic_intg}) and the expression (\ref{Wgn_def}) remains valid also when $r>N$. In this case the restriction 
on the length of the partition $l(\lambda)$ can be omitted. If so, and for $r>N$ the function $Wg^N(\sigma)$, as a function of $N$ 
will have poles at integers $-r+1,\cdots,r-1$. 
These poles will, however cancel after summation over all permutations in (\ref{basic_intg}).

For many applications in the lattice Quantum Chromodynamics (QCD) an extension 
of the integral ${\cal I}(r,s)$ to $SU(N)$ group is very important. 
Such integrals appear in the strong coupling expansion of the pure gauge action and are useful 
in the evaluation of the baryon spectrum. Integrals of this type appear during calculations 
of the dual representations of some spin models describing effective interaction between Polyakov loops 
\cite{spin_flux1, spin_flux2} and in principal chiral models \cite{su2pcm_dual}. 
Many one-link integrals in the strong-coupled LGT can be reduced to the computation of (\ref{integral_def}). 
These are one-link integrals with the staggered and the Wilson fermions as well as the one-link integrals 
of the scalar QCD. Most importantly, the knowledge of these integrals is necessary for the construction 
of the dual formulations of the full lattice QCD \cite{su2_dual, su3_dual, dual1}. 

In the paper \cite{dual1} we outlined a certain approach to the duality transformations based on the Taylor 
expansion of the Boltzmann weight both for $U(N)$ spin models and for $U(N)$ LGTs. Before presenting the details 
of our calculations and their development we think it would be useful to extend the integration methods to 
the special unitary group $SU(N)$. It will help in generalizing the approach of Ref. \cite{dual1} to lattice QCD. 
Actually, the generating functional for $SU(N)$ integrals ${\cal I}_N(r,s)$ has been calculated long ago (see the Appendix 
of the review \cite{lat_rev}). Recently, the case $s-r=N$ has been examined in some details in \cite{zuber}. 
In this paper we present our results for $SU(N)$ group integrals of the type (\ref{integral_def}) for arbitrary $r,s$. 
In deriving them we use the approach of Ref. \cite{novaes1}. 
In Section~2 we obtain and describe in details our results for $SU(N)$ integrals. 
As the simplest but important applications we compute various one-link and one-site 
integrals relevant for LGT in Section~3. The results and perspectives are discussed in Section~4. In Appendix we collected some 
formula used in computation of group integrals.

\section{$SU(N)$ polynomial integrals} 

In this section we calculate the following integral over $SU(N)$ group 
\begin{equation}
{\cal I}_N(r,s) \ = \  \int_{SU(N)}  dU \ \prod_{k=1}^{r} U_{i_k j_k} \prod_{n=1}^{s} U_{m_n l_n}^* \ . 
\label{integral_sundef}
\end{equation} 
Similar to $U(N)$ case this integral can be re-written as a sum of the invariant integrals. 
The product of the matrix elements in the integrand is presented in the form 
\begin{equation}
\prod_{k=1}^{r} U_{i_k j_k} \ = \ \frac{1}{r!} \ \prod_{k=1}^{r} \frac{\partial}{\partial A^*_{i_kj_k}} \ 
P_{1^r}(A^{\dagger}U) \ , 
\label{prod_matrix}
\end{equation}
where $P_{\tau}(X)$ is the power sum symmetric function (\ref{pwr_sum}). The integral becomes 
\begin{equation}
{\cal I}_N(r,s) \ = \  \frac{1}{r! s!} \ \prod_{k=1}^{r} \frac{\partial}{\partial A^*_{i_kj_k}} \ 
\prod_{n=1}^{s} \frac{\partial}{\partial B_{m_n l_n}} \ {\cal F} \ , 
\label{integral_sun1}
\end{equation}
where 
\begin{equation}
{\cal F} \ = \ \int_{SU(N)}  dU \ P_{1^r}(A^{\dagger}U) \ P_{1^s}(B U^{\dagger}) \ .
\label{integral_sun2}
\end{equation} 
Using Eq.(\ref{schur_pwrsum_inverse}) from Appendix the last integral can be brought to 
\begin{equation}
{\cal F} \ = \ \sum_{\lambda\vdash r} \ \chi_{\lambda}(1^r) \  \sum_{\mu\vdash s} \ \chi_{\mu}(1^s) \ 
\int_{SU(N)}  dU \ s_{\lambda}(A^{\dagger} U) \ s_{\mu}(B U^{\dagger}) \ . 
\label{integral_sun3}
\end{equation} 
We assume here the partitions $\lambda$ and $\mu$ are those described after Eq.(\ref{Wgn_def}). 
Consider first the following integral 
\begin{equation}
{\cal F}_0 \ = \ \int_{SU(N)}  dU \ s_{\lambda}(U) \ s_{\mu}(U^{\dagger}) \ . 
\label{integral_sun4}
\end{equation} 
Let $s_{\lambda}(U)=s_{\lambda}(u_1,\cdots,u_N)$, where $u_i$ are eigenvalues of $U$. Then 
\begin{equation}
s_{\lambda}(U) \ = \ \frac{{\rm det} \ u_i^{\lambda_j+N-j}}{{\rm det} \ u_i^{N-j}} \ .
\label{schur_det}
\end{equation}
Last two integrals can be written as integrals over $U(N)$ measure with additional constraint
\begin{equation}
 {\rm det} \ U \ = \ \prod_{i=1}^N \ u_i \ = \ 1 \ . 
\label{sun_det}
\end{equation}
Then, it is straightforward to prove 
\begin{equation}
{\cal F}_0 \ = \ \sum_{q=-\infty}^{\infty} \prod_{i=1}^N \ \delta_{\lambda_i-\mu_i+q,0} \ . 
\label{integral_sun5}
\end{equation}
Summation over $q$ enforces the constraint (\ref{sun_det}). We thus have 
\begin{equation}
\sum_{i=1}^N \ \left ( \mu_i - \lambda_i \right ) \ = \ s -r \ = \ N q \ , 
\label{triality_cond}
\end{equation}
{\it i.e.}, the difference $s-r$ must be multiple of $N$. This is, of course, a consequence 
of $Z(N)$ symmetry.  
Clearly, the same holds for the integral in (\ref{integral_sun3}), therefore we can write 
\begin{equation}
s_{\mu}(B U^{\dagger}) \ = \ s_{\lambda+q^N}(B U^{\dagger}) \ = \ s_{q^N}(B U^{\dagger})\ s_{\lambda}(B U^{\dagger}) 
\ = \ \left ( {\rm det} \ B U^{\dagger} \right )^q \ s_{\lambda}(B U^{\dagger}) \ , 
\label{schur_factor}
\end{equation}
where $\lambda+q^N=(\lambda_1+q,\cdots,\lambda_N+q)$ and $s_{q^N}(X)=s_{(q,\cdots,q)}(X)$. Last two equalities follow 
from the representation (\ref{schur_det}). Suppose that $s\geq r$.
Substituting last expression into (\ref{integral_sun3}) and performing integration we find 
\begin{equation}
{\cal F} \ = \ \sum_{q=0}^{\infty} \ \delta_{s-r,Nq} \left ( {\rm det} \ B \right )^q 
\ \sum_{\lambda\vdash r} \ d(\lambda) \  d(\lambda+q^N) \ 
\frac{s_{\lambda}(A^{\dagger} B)}{s_{\lambda}(1^N)} \ , 
\label{integral_sun6}
\end{equation} 
which results in the following intermediate result 
\begin{eqnarray}
{\cal I}_N(r,s) &=&  \sum_{q=0}^{\infty} \ \delta_{s-r,Nq} \ \frac{1}{r! (r+Nq)!} \ \ 
\sum_{\lambda\vdash r} \ \frac{d(\lambda) \  d(\lambda+q^N)}{s_{\lambda}(1^N)} \nonumber  \\  
&\times& \prod_{k=1}^{r} \frac{\partial}{\partial A^*_{i_kj_k}} \ 
\prod_{n=1}^{r+Nq} \frac{\partial}{\partial B_{m_n l_n}} \ 
s_{q^N}(B) \ s_{\lambda}(A^{\dagger} B) \ . 
\label{integral_sun7}
\end{eqnarray}
The equality $({\rm det} \ B)^q = s_{q^N}(B)$ was used here. 
To take all the derivatives one employs the relation between the Schur functions and the power sum symmetric 
functions (\ref{schur_pwrsum}). This gives for the second line of (\ref{integral_sun7}) 
\begin{eqnarray}
\frac{1}{r! (Nq)!} \  \sum_{\sigma\in S_r} \ \sum_{\tau\in S_{Nq}} \ \chi_{\lambda}(\sigma) \ 
\chi_{q^N}(\tau) \ \sum_{a_1,\cdots,a_r=1}^N \ \sum_{b_1,\cdots,b_r=1}^N \ \sum_{c_1,\cdots,c_{Nq}=1}^N \nonumber \\
\prod_{k=1}^{r} \frac{\partial}{\partial A^*_{i_kj_k}} \ \prod_{n=1}^{r+Nq} \frac{\partial}{\partial B_{m_n l_n}} \ 
\prod_{k=1}^{r} \ A^*_{b_k a_k} B_{b_k a_{\sigma (k)}}  \   \prod_{n=1}^{Nq} \ B_{c_n c_{\tau (n)}} \ .  
\label{derivativ}
\end{eqnarray}
Taking all derivatives and performing summations over matrix indices $a_i,b_i,c_i$ 
we finally obtain for $s\geq r$ 
\begin{eqnarray}
{\cal I}_N(r,s) &=&  \sum_{q=0}^{\infty} \ \delta_{s-r,Nq}  \ \sum_{\sigma\in S_r} \ Wg^{N,q}(\sigma)
\sum_{\rho\in S_{r+Nq}} \ \prod_{k=1}^r \ \delta_{i_k,m_{\rho(k)}} \delta_{j_{\sigma(k)},l_{\rho(k)}} 
\nonumber   \\ 
&\times& \frac{1}{(Nq)!} \ \sum_{\tau\in S_{Nq}} \ \chi_{q^N}(\tau) \ 
\prod_{k=1}^{Nq} \ \delta_{m_{\rho(\tau(k)+r)},l_{\rho(k+r)}} \ , 
\label{sunintegr_fin}
\end{eqnarray}
where we defined the $SU(N)$ Weingarten function as  
\begin{equation} 
Wg^{N,q}(\sigma) = \frac{1}{r!(r+Nq)!} \ \sum_{\lambda\vdash\ r} \ 
\frac{d(\lambda)d(\lambda + q^N)}{s_{\lambda}(1^N)} \ \chi_{\lambda}(\sigma) \ . 
\label{Wgn_SUN}
\end{equation} 
When $r=s$ one restores the result (\ref{unintg_def}). The extension to $r>s$ is trivial. 
The second line in (\ref{sunintegr_fin}) can be re-written in an equivalent form through completely 
anti-symmetric tensors with the help of identity 
\begin{eqnarray}
\sum_{\tau\in S_{Nq}} \ \chi_{q^N}(\tau) \ \prod_{k=1}^{Nq} \ \delta_{m_{\tau(k)},l_{k}} \ = \ 
\frac{1}{(N!)^q} \ \sum_{\sigma\in S_{Nq}} \ \epsilon_{m_{\sigma(1)}\cdots m_{\sigma(N)}} \cdots 
\nonumber   \\ 
\epsilon_{m_{\sigma((q-1)N+1)}\cdots m_{\sigma(qN)}} \ \epsilon_{l_{\sigma(1)}\cdots l_{\sigma(N)}} \cdots 
\epsilon_{l_{\sigma((q-1)N+1)}\cdots l_{\sigma(qN)}} \ . 
\label{epsilon_form}
\end{eqnarray}
While not so compact, this representation is more useful for practical calculations in LGT. 
The expressions (\ref{sunintegr_fin}), (\ref{Wgn_SUN}) and (\ref{epsilon_form}) are main result of this paper. 
They allow to evaluate many $SU(N)$ integrals and present the result in a compact form. 

Similar result for the generating function 
\begin{equation}
\int_{SU(N)}  dU \  \left ( \mbox{Tr}J U \right )^r \ \left ( \mbox{Tr}K U^{\dagger} \right )^s
\label{sun_genfunc}
\end{equation} 
has recently been derived in \cite{unger_lat18}. In this paper the authors define the $SU(N)$ Weingarten 
function as 
\begin{equation} 
\widetilde{Wg}^{N,q}(\sigma) = \frac{1}{(r!)^2} \ \sum_{\lambda\vdash\ r} \ 
\frac{d^2(\lambda)}{s_{\lambda}(1^{N+q})} \ \chi_{\lambda}(\sigma) \ . 
\label{Wgn_unger}
\end{equation} 
The relation between two definitions is given by 
\begin{equation} 
\prod_{k=0}^{N-1} \ \frac{k!}{(k+q)!}  \ \widetilde{Wg}^{N,q}(\sigma) \ =  \ Wg^{N,q}(\sigma) \ . 
\label{Wgn_relation}
\end{equation}
With this relation taken into account our result for the generating functional (\ref{sun_genfunc}) agree 
with that of \cite{unger_lat18}.

We finish this section with a conjecture about large-$N$ asymptotic behaviour of the $SU(N)$ Weingarten function. 
In case of the $U(N)$ group it is well-known and the leading term of the asymptotic expansion reads \cite{collins2} 
\begin{equation}
Wg^N(\sigma) \ = \ \frac{1}{N^{r}} \ 
\left ( \prod_{k} \ \delta_{\sigma(k),k} +  {\cal{O}} \left ( N^{-1}  \right ) \right ) \ , 
\label{WgUN_asymp}
\end{equation}
if $\sigma\in S_r$. The simple way to obtain the leading term for the $SU(N)$ group is to consider the integral 
(\ref{Q_def}) (see next section).  Then, one can readily find  
\begin{equation}
(r+Nq)! \ \sum_{\sigma\in S_r} \ Wg^{N,q}(\sigma)  N^{|\sigma|} \ = \ Q_N(r,s) \ .
\label{Wgn_sum}
\end{equation}
Here, $|\sigma|$ is a number of cycles in the permutation $\sigma$ and $Q_N(r,s)$ is given in Eq.(\ref{QSUN}). 
Taking asymptotics on the right-hand side of the last expression leads to the equality 
\begin{equation}
(r+Nq)! \ \sum_{\sigma\in S_r} \ Wg^{N,q}(\sigma) N^{|\sigma|} \ = \ \frac{(r+Nq)!}{(N!)^q} 
\left ( 1 + {\cal{O}} \left ( N^{-1}  \right )  \right) \ .
\label{Wgn_sum_asymp}
\end{equation}
If we suppose that, similar to $U(N)$ group, the leading contribution to the large-$N$ expansion comes from 
the identity permutation $\sigma(k)=k$ we obtain 
\begin{equation}
Wg^{N,q}(\sigma) \ = \ \frac{1}{N^{r}(N!)^q} \ \left ( \prod_{k} \ \delta_{\sigma(k),k}  +  
{\cal{O}} \left ( N^{-1}  \right )  \right ) \ . 
\label{WgSUN_asymp}
\end{equation}
Note, that $N^r (N!)^q\sim N^{r+Nq}=N^s$. 

\section{Applications}

In this section we apply the formula (\ref{sunintegr_fin}) for the evaluation of various $SU(N)$ integrals 
which often appear in some spin models and LGTs.

\subsection{Invariant integrals}

The simplest integrals which can be calculated with the help of (\ref{sunintegr_fin}) are those whose 
integrands depend only on the traces of the group elements  
\begin{equation} 
Q_N(r,s) = \int_G dU \ ({\rm Tr} U)^r \ ({\rm Tr} U^*)^s  \ , 
\label{Q_def}
\end{equation} 
where $G=U(N),SU(N)$. Such integrals appear in the effective spin models describing the interaction 
of the Polyakov loops in the finite-temperature LGTs. Writing the traces as 
\begin{equation}
({\rm Tr} U)^s \ = \ \sum_{i_1=1}^N \ \cdots   \sum_{i_s=1}^N \ 
\prod_{k=1}^s \ U_{i_ki_k}
\label{race_exp}
\end{equation}
one obtains the integral of the type (\ref{integral_def}) or (\ref{integral_sundef}). 
Using (\ref{basic_intg}) and (\ref{sunintegr_fin}) one finds after summation over group indices 
\begin{eqnarray} 
Q_N(r,s) \ = \  
\begin{cases}
\delta_{r,s } \  \sum_{\lambda \vdash s} \ d^2(\lambda) \ ,  \ G=U(N) \ ,  \\ 
\sum_{q=-\infty}^{\infty} \delta_{s-r,q N} \  
\sum_{\lambda \vdash {\rm min}(r,s)} \ d(\lambda) \ d(\lambda + |q|^N) \ , \ G=SU(N) \ . 
\end{cases}
\label{QSUN}
\end{eqnarray}
Of course, the simpler way to compute (\ref{Q_def}) is to use the expansion 
$({\rm Tr} U)^s=(u_1+u_2+\dots+u_N)^s=\sum_{\lambda \vdash s} d(\lambda)s_{\lambda}(U)$. 
Then, the result (\ref{QSUN}) follows from the orthogonality of the Schur functions. 

The next example is provided by the coefficients of the character expansion of an invariant function 
$H(x,y)$ 
\begin{equation}
C_{\lambda}(a,b) \ = \ \int_G dU \ s_{\lambda}(U) \ 
H\left ( \frac{a}{2}{\rm Tr} U, \frac{b}{2} {\rm Tr} U^{\dagger} \right ) \ . 
\label{char_coeff_def}
\end{equation}
For the most common case $H(x,y)=e^{x+y}$ the result is well known for many decades and reads for $G=SU(N)$
\begin{equation}
\label{char_coeff_res}
C_{\lambda}(a,b) \ = \ \left ( \frac{b}{a} \right )^{\frac{1}{2}|\lambda|} \ 
\sum_{m=-\infty}^\infty \ \left ( \frac{b}{a} \right )^{\frac{1}{2}N m}  \ 
{\rm det} I_{\lambda_i-i+j+m}(\sqrt{ab})_{1\leq i,j \leq N}\ , 
\end{equation} 
where $|\lambda|=\sum_i\lambda_i$. Only term with $m=0$ contributes for $U(N)$ group. 
An alternative expression for the coefficients $C_{\lambda}$ can be obtained with the help of (\ref{sunintegr_fin}). 
First, using the Taylor expansion of the function $H(x,y)$, the Eq.(\ref{char_coeff_def}) is written as 
\begin{equation}
\label{char_coeff_res1}
C_{\lambda}(a,b) \ = \ \sum_{r=0}^{\infty} \sum_{s=0}^{\infty} \ h(r,s) \ \frac{(a/2)^{r}}{r!} \ \frac{(b/2)^{s}}{s!} 
\int_G dU \ s_{\lambda}(U) \ ({\rm Tr} U)^r \ ({\rm Tr} U^{\dagger})^s \ , 
\end{equation} 
where $h(r,s) = {\left. \frac{\partial^{r+s}H(a,b)}{\partial a^{r}\partial b^{s}} \right |}_{a=b=0}$. 
The last integral can again be presented as an integral over the power sum symmetric functions following 
Eqs.(\ref{prod_matrix})-(\ref{integral_sun2}) and using the relations (\ref{schur_pwrsum}), (\ref{pwr_sum}). 
This leads to the computation of the integral of the form 
\begin{equation}
\int_G dU \ P_{\sigma}(U) \ P_{1^r}(A^{\dagger}U) \ P_{1^s}(B U^{\dagger}) \ , \ 
\sigma\in S_{|\lambda|} \ , \ |\lambda| = \sum_i\lambda_i \ . 
\label{}
\end{equation} 
Repeating all the calculations from the previous section for the last integral we end up with the following result 
\begin{equation}
\label{char_coeff_res2}
C_{\lambda}(a,b) \ = \ \sum_{r=0}^{\infty} \sum_{s=0}^{\infty} \ h(r,s)  \ \frac{(a/2)^{r}}{r!} \ \frac{(b/2)^{s}}{s!} \ 
\sum_{\mu\vdash r} \ \sum_{\nu\vdash s} \ C_{\lambda \ \mu}^{\nu}(G) \ d(\mu) \ d(\nu) \ .
\end{equation} 
The Littlewood-Richardson coefficients $C^{\nu}_{\lambda \ \mu}(G)$ appear in the final answer in the form 
\begin{equation}
C_{\lambda \ \mu}^{\nu}(U(N)) \ = \ \frac{1}{r!s!} \ \sum_{\sigma\in S_r} \ \sum_{\tau\in S_s} \ 
\chi_{\lambda}(\sigma) \ \chi_{\mu}(\tau) \ \chi_{\nu}([\sigma : \tau]) \ 
\label{LR_un}
\end{equation}
for $G=U(N)$ and 
\begin{equation}
C_{\lambda \ \mu}^{\nu}(SU(N)) \ = \ \sum_{q=0}^{\infty} \ (Nq)! \ \frac{d(\nu + |q|^N)}{s_{\nu}(1^N)} \ 
\sum_{\rho\vdash s+Nq} \ \frac{s_{\rho}(1^N)}{d(\rho)} \ C_{\nu \ Nq}^{\rho}(U(N)) \ 
C_{\lambda \ \mu}^{\rho}(U(N)) \ 
\label{LR_sun}
\end{equation}
for $G=SU(N)$. The permutation $[\sigma : \tau]\in S_{r+s}$ is defined as 
\begin{equation}
[\sigma : \tau](k) \ = \
\begin{cases}
 \sigma(k) \ , \ k\leq r \ , \\ 
 \tau(k-r)+r \ , \ r < k \leq r+s \ .
\end{cases}
\label{prod_def}
\end{equation}
The constraints on partitions $r+|\lambda|=s$ for $U(N)$ and $r+|\lambda|=s+Nq$ for $SU(N)$ are evident 
from the Littlewood-Richardson coefficients. The expression (\ref{char_coeff_res2}) can be further simplified 
using the summation formula 
\begin{equation}
\sum_{\mu\vdash r} \ C_{\lambda \ \mu}^{\nu}(U(N)) \ d(\mu) \ = \ d\left ( \nu/\lambda  \right ) \ .
\label{Cd_sum}
\end{equation} 
Here, $d\left ( \nu/\lambda  \right )$ is the dimension of a skew representation defined by a corresponding 
skew Young diagram 
\begin{equation}
d\left ( \nu/\lambda  \right ) \ = \ \frac{1}{|\lambda|!} \sum_{\sigma\in S_{|\lambda|}} \ \chi_{\nu}([I : \sigma]) \ 
\chi_{\lambda}(\sigma) \ .  
\label{skew_d}
\end{equation}
Then, {\it e.g.} for $U(N)$ we get a simple expression 
\begin{equation}
\label{char_coeff_res3}
C_{\lambda}(a,b) \ = \ \sum_{r=0}^{\infty} \ h(r,r+|\lambda|) \ 
\frac{(ab/4)^{r}}{r!} \ \frac{(b/2)^{|\lambda|}}{(r+|\lambda|)!} \ 
\sum_{\nu\vdash r+|\lambda|} \ d\left ( \nu/\lambda  \right ) \ d(\nu)
\end{equation} 
and similar formula can be written down for $SU(N)$ if one uses Eqs.(\ref{char_coeff_res2}), (\ref{LR_sun}).

\subsection{One-link integrals}

Consider the following integral over $G=U(N),SU(N)$ with arbitrary $N\times N$ matrices $A,B$ 
\begin{equation}
Z_G(A,B) \ = \ \int_G dU \ \exp \left [ \frac{\beta}{2} \left ( {\rm Tr} A U 
+ {\rm Tr} B U^{\dagger}  \right )   \right  ] \ . 
\label{one_linkint_def}
\end{equation} 
Expanding the integrand in the Taylor series one gets 
\begin{equation}
Z_G(A,B) \ = \ \sum_{r=0}^{\infty}\sum_{s=0}^{\infty} \ \frac{(\beta/2)^{r+s}}{r!s!} \ 
\int_G \ dU \left ( {\rm Tr} AU \right )^r \ \left ( {\rm Tr} BU^{\dagger} \right )^s \ . 
\label{one_linkint_exp1}
\end{equation} 
This can be presented as 
\begin{equation}
Z_G(A,B) =  \sum_{r=0}^{\infty}\sum_{s=0}^{\infty} \ \frac{(\beta/2)^{r+s}}{r!s!} 
\sum_{i_1\cdots i_r} \ \sum_{j_1\cdots j_r} \ \sum_{m_1\cdots m_s} \ \sum_{l_1\cdots l_s} 
\prod_{k=1}^r A_{{i_k j_k}} \ \prod_{n=1}^s B_{l_n m_n} \  {\cal I}_N(r,s) \ . 
\label{one_linkint_exp2}
\end{equation} 
Using the result of the previous section (\ref{sunintegr_fin}) for the group integral 
we can calculate all summations over group indices and over all permutations. All necessary formulae 
to do this are given in the Appendix. We find for $G=SU(N)$
\begin{eqnarray}
Z_{SU(N)}(A,B)  =  \sum_{s=0}^{\infty} \sum_{q=-\infty}^{\infty} \ 
\frac{1}{s!} \left ( \frac{\beta}{2} \right )^{2s+N|q|} \ \left ( \mbox{det} C \right )^{|q|} 
\sum_{\sigma \in S_s} \  Wg^{N,q}(\sigma) P_{\sigma} (A B) \ . 
\label{sun_1link_res_Wgn}
\end{eqnarray}
With the help of (\ref{char_orthog}) and (\ref{schur_pwrsum_inverse}) this can be re-written as 
\begin{eqnarray}
Z_{SU(N)}(A,B)  =  \sum_{s=0}^{\infty} \sum_{q=-\infty}^{\infty} \ \left ( \frac{\beta}{2} \right )^{2s+N|q|} \ 
\sum_{\lambda \vdash s} \ \frac{d(\lambda)d(\lambda + |q|^N)}{s!(s+N|q|)!} \ \frac{s_{\lambda}(A B)}{s_{\lambda}(1^N)} 
\left ( \mbox{det} C \right )^{|q|} , 
\label{sun_1link_res}
\end{eqnarray}
where $C=A$ if $q>0$ and $C=B$ if $q<0$. Equivalent form of (\ref{sun_1link_res}) reads 
\begin{eqnarray}
Z_{SU(N)}(A,B) &=&  \sum_{q=-\infty}^{\infty} \ \sum_{\lambda_1\geq\lambda_2\geq\cdots\geq\lambda_N\geq 0} \ 
\left ( \frac{\beta}{2} \right )^{2\sum_i\lambda_i+N|q|} \ 
\frac{d(\lambda)d(\lambda + |q|^N)}{\left ( \sum_i\lambda_i \right )!\left ( \sum_i\lambda_i+N|q| \right )!} \nonumber   \\ 
&\times& \ \frac{s_{\lambda}(A B)}{s_{\lambda}(1^N)} \ \left ( \mbox{det} C \right )^{|q|} \ . 
\label{sun_1link_fin}
\end{eqnarray}
When $B=0$ the last formula reduces to 
\begin{eqnarray}
Z_{SU(N)}(A,0)  =  \sum_{q=0}^{\infty} \ \left ( \frac{\beta}{2} \right )^{Nq} \
\left ( \mbox{det} A \right )^q \ 
\prod_{k=0}^{N-1} \ \frac{k!}{(q+k)!} \ . 
\label{sun_1link_B0}
\end{eqnarray}
This result coincides with \cite{wipf} for $A=a I$, where $I$ is a unit matrix.
For $U(N)$ group (\ref{sun_1link_fin}) simplifies to 
\begin{eqnarray}
Z_{U(N)}(A,B) \ = \ \sum_{\lambda_1\geq\lambda_2\geq\cdots\geq\lambda_N\geq 0} \ 
\left ( \frac{\beta}{2} \right )^{2\sum_i\lambda_i} \ 
\frac{d^2(\lambda)}{\left ( ( \sum_i\lambda_i )! \right )^2} \ 
\frac{s_{\lambda}(A B)}{s_{\lambda}(1^N)} \ . 
\label{un_1link_res}
\end{eqnarray} 
Integrals of the more general form 
\begin{equation}
Z_G(A,B,C_k,D_l) \ = \ \int_G dU \ \prod_k \mbox{Tr} C_k U \ \prod_l \mbox{Tr} D_l U^{\dagger}
\exp \left [ \frac{\beta}{2} \left ( {\rm Tr} A U 
+ {\rm Tr} B U^{\dagger}  \right )   \right  ]  
\label{one_linkint_gen}
\end{equation} 
can be evaluated by differentiating the basic integral (\ref{one_linkint_def}), for example 
\begin{equation}
Z_G(A,B,C,D) \ = \ {\left. \frac{\partial^{2}}{\partial h_1 \partial h_2} \ Z_G(A+h_1 C , B+h_2 D) \right |}_{h_1=h_2=0} \ . 
\label{one_linkint_gen_res}
\end{equation} 
The final result is quite cumbersome and will not be given here.

\subsection{Staggered fermions} 

Formulae (\ref{sun_1link_res_Wgn})-(\ref{un_1link_res}) can be further specified for many particular cases. 
As an important example, let us consider the partition function with staggered fermions. 
For $N_f$ flavours of the staggered fermions the matrices $A$ and $B$ from (\ref{one_linkint_def}) read 
\begin{equation}
A^{ij} \ = \  \eta_{\nu}(x) \ y_{\nu} \ \sum_{f=1}^{N_f} \ 
\bar{\psi}^i_f(x) \psi^j_f(x+e_{\nu}) \ ,  
\label{A_stagferm}
\end{equation}
\begin{equation}
B^{ij} \ = \ - \eta_{\nu}(x) \ y_{\nu}^{-1} \ \sum_{f=1}^{N_f} \ 
\bar{\psi}^i_f(x+e_{\nu}) \psi^j_f(x) \ . 
\label{B_stagferm}
\end{equation}
$\eta_{\nu}(x)$ is given by 
\begin{equation} 
\eta_{0}(x) \ = \ 1 \ ; \ 
\eta_{\nu}(x) \ = \  \xi \ (-1)^{x_0+x_{1}+ \cdots +x_{\nu-1}} \ , \ \nu = 1,\cdots , d \ .
\label{KSferm}
\end{equation}
$\xi=a_t/a_s$ is the lattice anisotropy and the chemical potential $\mu$ is introduced via $y_{\nu}$ 
\begin{equation}
y_{\nu} \ = \ 
\begin{cases}
\exp (\mu), \ \nu = 0 \ , \ \mu = a_t \mu_q  \\ 
1, \  \nu=1,\cdots , d \ .
\end{cases}
\label{chem_pot}
\end{equation}
We define the composite meson fields as 
\begin{equation} 
\sigma_{f f^{\prime}}(x) \ = \ \sum_{i=1}^N \ \bar{\psi}^{i}_f(x)\psi^i_{f^{\prime}}(x) 
\label{meson_comp} 
\end{equation} 
and the composite (anti-)baryon fields as 
\begin{eqnarray} 
\label{baryon_comp} 
&&B_{f_1...f_{N}}(x) \ = \ \frac{1}{N!} \ \sum_{i_1...i_{N}=1}^{N} \ 
\epsilon_{i_1...i_{N}} \ \psi^{i_1}_{f_1}(x) \cdots \psi^{i_{N}}_{f_{N}}(x) \ , \\
&&\bar{B}_{f_1...f_{N}}(x) \ = \ \frac{1}{N!} \ \sum_{i_1...i_{N}=1}^{N} \ \epsilon_{i_1...i_{N}} \ 
\bar{\psi}^{i_{N}}_{f_1}(x) \cdots \bar{\psi}^{i_1}_{f_{N}}(x) \ .
\label{antibaryon_comp} 
\end{eqnarray} 
Consider first the $U(N)$ model (\ref{un_1link_res}). Using representation (\ref{schur_repr1}) for the Schur functions  
and definitions (\ref{A_stagferm}), (\ref{B_stagferm}) it is easy to prove that 
\begin{equation}
{\rm Tr} (AB)^j \ = \ \left ( \eta_{\nu}(x) \right )^{2j} \ (-1)^{j+1} \ {\rm Tr} \ \Sigma^j \ , 
\label{AB_trace}
\end{equation}
where we introduced matrix 
\begin{equation}
\Sigma_{f_1f_2} \ = \ \sum_{f=1}^{N_f} \sigma_{f_1f}(x)\sigma_{ff_2}(x+e_{\nu}) \ . 
\label{sigma_def}
\end{equation}
Substituting this result into (\ref{un_1link_res}) and using $\sum i\tau_i=s$, we obtain ($\beta=1$) 
\begin{eqnarray}
Z_{U(N)}(A,B) \ = \ \sum_{s=0}^{N N_f} \ \left ( \frac{\eta_{\nu}(x)}{2} \right )^{2s} \ \sum_{\lambda\vdash\ s} \ 
\frac{d^2(\lambda)}{(s!)^2} \ \frac{s_{\lambda^{\prime}}(\Sigma)}{s_{\lambda}(1^N)}  \ . 
\label{un_1link_stagferm}
\end{eqnarray}
Here, $\lambda^{\prime}$ is a representation dual to $\lambda$. The dual representation is defined by exchanging 
raws and columns in the corresponding Young diagram, {\it i.e.} $\lambda^{\prime}_i=\sum_j \mathbf{1}_{\lambda_j\geq i}$. 
In order to extend this result to $SU(N)$ we need to calculate determinants of $A$ and $B$ matrices. They are 
\begin{eqnarray}
\label{DetA}
\mbox{det} A  = (-1)^{\frac{N(N-1)}{2}} N! (\eta_{\nu}(x) \ y_{\nu})^N \sum_{f_i = 1}^{N_f} 
\bar{B}_{f_1...f_{N}}(x) B_{f_1...f_{N}}(x+e_{\nu})  \ , \\
\mbox{det} B  = (-1)^{\frac{N(N-1)}{2}} N! (- \eta_{\nu}(x) \ y_{\nu}^{-1})^N \sum_{f_i = 1}^{N_f} 
\bar{B}_{f_1...f_{N}}(x+e_{\nu}) B_{f_1...f_{N}}(x) \ .
\label{DetB}
\end{eqnarray}
The factor $(-1)^{\frac{N(N-1)}{2}}$ appears due to $\sum_{i = 0}^{N-1} i = \frac{N(N-1)}{2}$ commutations 
of the fermion fields to gather them into baryons. Combining these expressions with (\ref{un_1link_stagferm}) and 
(\ref{sun_1link_res}) we can write down the partition function with an arbitrary number of staggered fermion flavours as 
\begin{eqnarray}
Z_{SU(N)}(A,B)  &=&  Z_{U(N)}(A,B)  + 
\sum_{q=1}^{N_f} \ (-1)^{\frac{N(N-1)}{2}q} (N!)^q \ \sum_{s=0}^{N (N_f-q)} \ \left ( \frac{\eta_{\nu}(x)}{2} \right )^{2s+Nq} 
\nonumber   \\  
&\times& \sum_{\lambda \vdash s} \ \frac{d(\lambda)d(\lambda + q^N)}{s!(s+N q)!} 
\frac{s_{\lambda^{\prime}}(\Sigma)}{s_{\lambda}(1^N)}  \ 
\left [ y_{\nu}^{Nq} Q^q +  (-1)^{Nq} y_{\nu}^{-Nq} \bar{Q}^q  \right ] \ , 
\label{PF_stagferm}
\end{eqnarray}
where notations have been used 
\begin{eqnarray}
Q \ = \ \sum_{f_i = 1}^{N_f} \ \bar{B}_{f_1...f_{N}}(x) B_{f_1...f_{N}}(x+e_{\nu}) \ , \nonumber \\ 
\bar{Q} \ = \ \sum_{f_i = 1}^{N_f}  \bar{B}_{f_1...f_{N}}(x+e_{\nu}) B_{f_1...f_{N}}(x) \ . 
\label{Qbaryon}
\end{eqnarray}
As the simplest case, consider one-flavour system. Then, the matrix $\Sigma$ becomes a scalar 
$\Sigma=\sigma(x)\sigma(x+e_{\nu})$ and the summation over $\lambda$ consists of one term, namely 
$\lambda^{\prime}=(s,0,\cdots,0)$ (or $\lambda_i=1, i\in [1,s]$ and zero otherwise). 
This, together with (\ref{schur_reduction}), yields 
\begin{eqnarray}
Z_{SU(N)}(A,B) \ = \ \sum_{s=0}^{N} \ \left ( \frac{\eta_{\nu}(x)}{2} \right )^{2s} \ 
\frac{(N-s)!}{s!N!} \ \left ( \sigma(x)\sigma(x+e_{\nu})  \right )^s \nonumber   \\ 
+ (-1)^{\frac{N(N-1)}{2}}  \left ( \frac{\eta_{\nu}(x)}{2} \right )^N \ 
\left [ y_{\nu}^N \bar{B}(x) B(x+e_{\nu}) + (-1)^N y_{\nu}^{-N} \bar{B}(x+e_{\nu}) B(x)  \right ] \ . 
\label{un_1link_stagferm_nf1}
\end{eqnarray}
This agrees with the well-known result of Ref.\cite{rossi}. 
Similar method of computing one-link integral with one flavour of the staggered fermions 
has been used in \cite{strcpl_qcd}. Slightly more complicated is the case of two staggered flavours, $N_f=2$. 
Parameterizing  $\lambda^\prime$ as $\lambda^\prime = (u_1, u_2)$, $u_1 + u_2 = s$, $N \geqslant u_1 \geqslant u_2 \geqslant 0$ 
and utilizing the property (\ref{schur_reduction}) of the Schur functions 
we present result for the two-flavour partition function as 
\begin{equation}
Z_{SU(N)}(A,B) \ = \ Z_0 + Z_1 + Z_2 \ , 
\label{2flavorPF}
\end{equation}
\begin{eqnarray}
Z_0 \ &=& \ \sum_{u_1 \geqslant u_2 \geqslant 0}^N 
 \left( \frac{\eta_{\nu}(x)}{2} \right)^{2 (u_1 + u_2)} 
 \frac{(u_1 - u_2 + 1) (N - u_1)! (N - u_2 + 1)!}{(u_1 + 1)! u_2! N! (N + 1)!} \nonumber \\
 &\times& \sum_{i = 0}^{\left[ \frac{u_1 - u_2} {2} \right]} 
 C_{u_1 - u_2 - i}^{i} (-1)^i (\mbox{det}\Sigma)^{u_2 + i} (\mbox{Tr} \Sigma )^{u_1 - u_2 - 2i} \ ,
\label{2flavorPF0}
\end{eqnarray}
\begin{eqnarray}
\label{2flavorPF1}
&&Z_1  = \sum_{u_1 \geqslant u_2 \geqslant 0}^N 
 \left( \frac{\eta_{\nu}(x)}{2} \right)^{2 (u_1 + u_2) +  N} 
 \frac{(u_1 - u_2 + 1) (N - u_1 + 1)! (N - u_2 + 2)!}{(u_1 + 1)! u_2! (N + 1)! (N + 2)!}  \\
&&(-1)^{\frac{N(N-1)}{2}} [ y_{\nu}^N Q + (-1)^N y_{\nu}^{-N} \bar{Q} ] \sum_{i = 0}^{\left[ \frac{u_1 - u_2} {2} \right]} 
 C_{u_1 - u_2 - i}^{i} (-1)^i (\mbox{det}\Sigma)^{u_2 + i} (\mbox{Tr} \Sigma )^{u_1 - u_2 - 2i} , \nonumber
\end{eqnarray}
\begin{eqnarray}
Z_2 \ = \ \left ( \frac{\eta_{\nu}(x)}{2} \right )^{2 N} \ 
\frac{1}{(N+1)} \ \left [ y_{\nu}^{2N} Q^2 + y_{\nu}^{-2N} \bar{Q}^2 \right ] \ . 
\label{2flavorPF2}
\end{eqnarray}
Here, $C^i_k$ are the binomial coefficients. To calculate the Schur functions we have used their relations 
with the complete symmetric functions $h_k$ and the elementary symmetric functions $e_k$, 
(\ref{ef_hf})-(\ref{ef_hf_schur}).

\subsection{Reduced principal chiral model} 

Our last example is the so-called reduced principal chiral model whose partition function reads 
\begin{equation}
Z \ = \ \int_G dU \ \exp \left [ \beta \sum_{\mu=1}^d  {\rm Re} {\rm Tr} U \Gamma_{\mu}  
U^{\dagger} \Gamma_{\mu}^{\dagger} \right  ] \ .
\label{pcm_reduced}
\end{equation} 
The matrices $\Gamma_{\mu}\in SU(N)$ and satisfy certain commutation relations exact form of which is not 
important here. It is expected that the reduced model gives an exact solution for the principal chiral model 
in the large-$N$ limit (see, for instance (\cite{vicari}) and refs. therein). Here we discuss the group integration 
in (\ref{pcm_reduced}) using the method of the Weingarten functions. For simplicity, we restrict ourself to the two-dimensional 
case. Expanding the integrand into two character series and using the relation (\ref{schur_pwrsum}) again one finds  
\begin{equation}
Z \ = \ \sum_{\lambda_1} \ \sum_{\lambda_2} \ \frac{C_{\lambda_1}(\beta) \ C_{\lambda_2}(\beta)}{|\lambda_1|!|\lambda_2|!} \ 
\sum_{\sigma_1\in S_{|\lambda_1|}} \ \sum_{\sigma_2\in S_{|\lambda_2|}} \ \chi_{\lambda_1}(\sigma_1) \ \chi_{\lambda_2}(\sigma_2) 
J_{\sigma_1,\sigma_2}(\{ \Gamma_{\mu} \}) \ , 
\label{pcm_reduced_exp}
\end{equation}
where $C_{\lambda_i}(\beta)$ are given by either (\ref{char_coeff_res}) or (\ref{char_coeff_res3}) and the resulting group 
integral takes the form 
\begin{equation}
J_{\sigma_1,\sigma_2}(\{ \Gamma_{\mu} \}) \ = \ \int_G dU \ P_{\sigma_1}(U\Gamma_1U^{\dagger}\Gamma_1^{\dagger}) \ 
P_{\sigma_2}(U\Gamma_2U^{\dagger}\Gamma_2^{\dagger}) \ . 
\label{Jss_def}
\end{equation}
Applying Eq.(\ref{pwr_sum}) one sees the last integral is of a type (\ref{integral_sundef}), therefore 
the equation (\ref{sunintegr_fin}) can be used (it is sufficient to take only the term with $q=0$ in the limit of large $N$) 
and gives after partial summation over group indices 
\begin{equation}
J_{\sigma_1,\sigma_2}(\{ \Gamma_{\mu} \}) \ = \ \sum_{\tau,\rho\in S_{r_1+r_2}} \ 
Wg^N(\tau^{-1}\rho) \ \overline{P}_{\tau,\tau}(\{ \Gamma_{\mu} \}) \ 
\overline{P}_{\rho\sigma_1,\rho\sigma_2}(\{ \Gamma_{\mu}^{\dagger} \}) \ . 
\label{Jss_res}
\end{equation} 
We introduced here the following invariant function 
\begin{equation}
\overline{P}_{\tau,\pi}(\{ \Gamma_{\mu} \})  \ = \ \sum_{i_1,\cdots,i_{r_1+r_2}} \ 
\prod_{k=1}^{r_1} \Gamma_1^{i_{\tau(k)},i_k} \ 
\prod_{k=1}^{r_2} \Gamma_2^{i_{\pi(r_1+k)},i_{r_1+k}} \ . 
\label{P_inv}
\end{equation}
The right-hand side of the last equation can be expressed in terms of traces of products of matrices $\Gamma_{\mu}$. 
Detailed investigation will be published elsewhere. Here we would like to emphasize that we expect certain simplifications 
in the large-$N$ limit which is the only one relevant here. 
Indeed, taking the asymptotics of the Weingarten function (\ref{WgUN_asymp}) 
in  Eq.(\ref{Jss_res}) we obtain 
\begin{equation}
J_{\sigma_1,\sigma_2}(\{ \Gamma_{\mu} \}) \ = \ \sum_{\rho\in S_{r_1+r_2}} \ 
\frac{1}{N^{r_1+r_2}} \ \overline{P}_{\rho,\rho}(\{ \Gamma_{\mu} \}) \ 
\overline{P}_{\rho\sigma_1,\rho\sigma_2}(\{ \Gamma_{\mu}^{\dagger} \}) \ . 
\label{Jss_res1}
\end{equation}

\section{Summary and perspectives} 

In this article we have calculated certain integrals over $SU(N)$ group. 
The basic integral (\ref{integral_sundef}) is of polynomial type whose integrand includes product
of an arbitrary number of the group matrices and their conjugates in the fundamental representation. 
The result of the integration is expressed through the summation over permutations of the group indices 
which are contracted via products of the Kronecker deltas and the totally anti-symmetric tensors, 
Eqs.(\ref{sunintegr_fin}) and (\ref{epsilon_form}). The weight is given by the $SU(N)$ Weingarten 
function (\ref{Wgn_SUN}). We have considered several applications of this general result. 
In particular, we have calculated the integrals of powers of traces of the group elements, 
the coefficients of the character expansion of the invariant function of an arbitrary form and 
the general one-link integral appearing in LGT. In the latter case, the result was used to evaluate 
one-link integrals with arbitrary number of staggered fermion flavours. Also, we applied our approach 
for the computation of the integrals in the expression for the partition function of the reduced principal chiral model. 

The method of the Weingarten functions is a powerful and very general tool which can be used both for the 
derivation of alternative representations of known integrals and for calculation of new and 
more complex ones appearing in lattice QCD and spin models. 
Some further applications of this method may include one-link integrals with $N_f$ flavours of the Wilson 
fermions, the scalar lattice QCD in the strong coupling region, $SU(N)$ principal chiral model at finite $N$. 
Interesting is an extension of 3.3 to the Eguchi-Kawai model and its twisted version. 

The main goal of this article was to give a mathematical background to our computations of the dual representations 
of 1) the effective Polyakov loop spin models and 2) pure gauge LGT and lattice QCD with the staggered fermions, 
as outlined in \cite{dual1}. Details of these calculations will be reported elsewhere. 
Also, an interesting and important direction is to explore the large-$N$ expansion of $SU(N)$ models in frameworks of  
the present method. This can be done by using the asymptotic expansion of the Weingarten function (\ref{WgSUN_asymp}) 
and is planned for future investigations.

\section*{Appendix} 

Let $\lambda=(\lambda_1,\lambda_2,\cdots,\lambda_N)$ be a partition $\lambda\vdash\ r$, 
$\lambda_1\geq\lambda_2\geq \cdots \geq \lambda_N\geq 0$ and $\sum_{i=1}^N\lambda_i=r$. 
$\chi_{\lambda}(\sigma)$ denotes a character of $\sigma \in S_r$ 
in representation $\lambda$. $d(\lambda)=\chi_{\lambda}(1)$ is the dimension of the representation $\lambda$. 
The Schur function $s_{\lambda}(X)$ is a character of the unitary group $G$, thus $s_{\lambda}(1^N)$ is 
the  dimension of the irreducible representation $\lambda$ of $G$. If $l(\lambda)$ is the length of the partition $\lambda$, 
{\it i.e.}, the number of non-vanishing parts $\lambda_i$, then 
\begin{equation}
d(\lambda) \ = \ r! \ 
\frac{\prod_{1\leq i<j \leq l(\lambda)}(\lambda_i-\lambda_j+j-i)}{\prod_{i=1}^{l(\lambda)}(\lambda_i+l(\lambda)-i)!} \ , 
\label{dim_l}
\end{equation}
\begin{equation}
s_{\lambda}(1^N) \ = \ \frac{\prod_{1\leq i<j \leq N}(\lambda_i-\lambda_j+j-i)}{\prod_{i=1}^N( N-i)!} \ . 
\label{G_dim}
\end{equation} 
Below we list some formulae used in evaluating the group integrals. 
Orthogonality relation for the characters of the permutation group is written as 
\begin{equation} 
\sum_{\tau\in S_r} \ \chi_{\mu}(\tau) \chi_{\lambda}(\tau \sigma) \ = \ 
r! \ \delta_{\mu,\lambda} \ \frac{\chi_{\lambda}(\sigma)}{d(\lambda)} \ .
\label{char_orthog}
\end{equation}
Relation between Schur functions and power sum symmetric functions of matrix argument is given by 
\begin{equation} 
s_{\lambda}(X) \ = \ \frac{1}{r!} \ \sum_{\tau\in S_r} \ \chi_{\lambda}(\tau) P_{\tau}(X) \ , 
\label{schur_pwrsum}
\end{equation} 
where $X$ has the dimension $N$ and 
\begin{equation}
P_{\tau}(X) \ = \ \sum_{i_1i_2\cdots i_r}^N \ \prod_{k=1}^r \ X_{i_k i_{\tau (k)}} \ . 
\label{pwr_sum}
\end{equation} 
Inverse relation reads 
\begin{equation} 
P_{\tau}(X) \ = \ \sum_{\mu\vdash r} \ \chi_{\mu}(\tau) s_{\mu}(X) \ .  
\label{schur_pwrsum_inverse}
\end{equation} 
The power sum symmetric function of the unit matrix equals 
\begin{equation} 
P_{\tau}(I) \ = \ N^{|\tau|} \ , 
\label{pwrsum_unitmatr}
\end{equation} 
where $|\tau|$ is the number of cycles in the permutation $\tau$. 
We mention the following formula  ($1 \leq i \leq N$)
\begin{equation}
\sum_{j_1\cdots j_r}^N \ \sum_{m_1\cdots m_r}^N \ \prod_{k=1}^r \  
\delta_{j_k,m_{\tau (k)}} \ \delta_{j_k,m_{\sigma (k)}} \ y_{j_k} y_{m_k} \ = \ 
P_{\tau^{-1}\sigma}(x) \ , \ x_i=y_i^2  \ . 
\label{pwr_sum_add}
\end{equation} 

Given the complete symmetric functions $h_k$ and the elementary symmetric functions $e_k$ 
in $m$ variables $x_1,\cdots ,x_m$ 
\begin{eqnarray}
h_k \ = \ \sum_{1\leq n_1\leq\cdots\leq n_k \leq m} \ x_{n_1} \cdots x_{n_k} \ , \ 
e_k \ = \ \sum_{1\leq n_1 < \cdots < n_k \leq m} \ x_{n_1} \cdots x_{n_k} \ , 
\label{ef_hf}
\end{eqnarray} 
the Schur functions can be computed with the help of identities 
\begin{eqnarray} 
s_{\lambda}(X) \ = \ s_{(\lambda_1,\cdots,\lambda_m )}(x_1,\cdots,x_m) \ = \ 
\mbox{det} \left ( h_{\lambda_i-i+j}  \right )_{1\leq i,j \leq m} \ , \nonumber  \\ 
s_{\lambda}(X) \ = \ s_{(\lambda_1,\cdots,\lambda_m )}(x_1,\cdots,x_m) \ = \ 
\mbox{det} \left ( e_{\lambda^{\prime}_i-i+j}  \right )_{1\leq i,j \leq m} \ , 
\label{ef_hf_schur}
\end{eqnarray}
where $\lambda^{\prime}$ is a partition dual to $\lambda$ and the following rule is understood 
\begin{equation}
s_{(\lambda_1,\cdots,\lambda_n)}(x_1,\cdots,x_{n-1},0) \ = \   
\begin{cases}
0, \ {\rm {if}} \ \lambda_n \ne 0  \ \ , \\ 
s_{(\lambda_1,\cdots,\lambda_{n-1})}(x_1,\cdots,x_{n-1}), \ {\rm {if}} \ \lambda_n=0 \ . 
\end{cases}
\label{schur_reduction}
\end{equation} 
Another expression for the Schur function used in the text reads 
\begin{equation} 
s_{\lambda}(X) \ = \ \sum_{\tau_1,...,\tau_s; \sum i\tau_i=s} \ \chi_{\lambda}(\tau) 
\prod_{j=1}^s \ \frac{1}{\tau_j! j^{\tau_j}}  \  \left [  {\rm Tr} (X)^j \right ]^{\tau_j} 
\label{schur_repr1}
\end{equation}
with $\tau$ being a permutation such that the number of cylcles of length $j$ in $\tau$ is $\tau_j$.

\vspace{0.5cm}

{\bf \large Acknowledgements}

\vspace{0.2cm}

V.C. acknowledges financial support from the INFN HPC{\_}HTC
and NPQCD projects.


\vspace{0.5cm}



\begin{thebibliography}{99} 

%
\bibitem{creutz} M.~Creutz, J. Math. Phys. {\bf 19} (1978) 2043. 
%
\bibitem{lat_rev} J.-M.~Drouffe, J.-B.~Zuber, Phys. Rep. {\bf 102}  (1983) 1. 
%
\bibitem{weingarten} D.~Weingarten, J.Math.Phys. {\bf 19} (1978) 999.
%
\bibitem{bars_green} I.~Bars, F.~Green, Phys. Rev. {\bf D20} (1979) 3311. 
%
\bibitem{bars} I.~Bars, J. Math. Phys. {\bf 21} (1980) 2678.
%
\bibitem{samuel_un} S.~Samuel, J. Math. Phys. {\bf 21} (1980) 2695. 
%
\bibitem{eriksson} K.E.~Eriksson, N.~Svartholm, B.S.~Skagerstam, 
J. Math. Phys. {\bf 22} (1981) 2276.
%
\bibitem{brower} R.C.~Brower, P.~Rossi, C.I.~Tan, Phys.Rev. {\bf D23} (1981) 942. 
%
\bibitem{balantekin} A.B.~Balantekin, Phys. Rev. {\bf D62} (2000) 085017. 
%
\bibitem{kontsevich_mod} A.~Mironov, A.~Morozov, G.~W.~Semenoff, Int. J. Mod. Phys. 
{\bf A11} (1996) 5031. 
%
\bibitem{moerbeke1} P. van Moerbeke ed., (2001). Integrable lattices: random matrices and random permutations. 
In: Random Matrix Models and Their Applications. Cambridge University Press, pp. 321-406; arXiv:math/9912143 [math.CO] (1999).
%
\bibitem{moerbeke2} M.~Adler, P. van Moerbeke, Comm. Pure Appl. Math. {\bf 54} (2001) 153. 
%
\bibitem{orlov1} A.~Yu.~Orlov, Tau functions and matrix integrals, arXiv:math-ph/0210012v3 (2002). 
%
\bibitem{orlov2} J.~W. van de Leur, A.~Yu.~Orlov, J. Phys. A: Math. Theor. {\bf 51} (2018) 025208. 
%
\bibitem{orlov3} A.~Yu.~Orlov, Matrix integrals and Hurwitz numbers, 
arXiv:1701.02296v4 [math-ph] (2017). 
%
\bibitem{wipf} S.~Uhlmann, R.~Meinel, A.~Wipf, J. Phys. A: Math. Theor. {\bf 40} (2007) 4367. 
%
\bibitem{carlsson} J.~Carlsson, Integrals over SU(N), arXiv:0802.3409 [hep-lat]. 
%
\bibitem{collins1} B.~Collins, Int.Math.Res.Not. {\bf 17} (2003) 952. 
%
\bibitem{collins2} B.~Collins, P.~\'Sniady, Commun.Math.Phys. {\bf 264} (2006) 773.  
%
\bibitem{novak} J.~Novak, Complete homogeneous symmetric polynomials in Jucys-Murphy elements 
and the Weingarten function, arXiv:0811.3595 [math.CO]. 
%
\bibitem{zinn-justin} P.~Zinn-Justin, Lett.Math.Phys. {\bf 91} (2) (2010) 119. 
%
\bibitem{novaes1} M.~Novaes, Elementary derivation of Weingarten functions of classical Lie groups, 
arXiv:1406.2182v2 [math-ph]. 
%
\bibitem{novaes2} M.~Novaes, J. Phys. A: Math. Theor. {\bf 50} (2016) 075201.
%
\bibitem{collins3} B.~Collins, S.~Matsumoto, Lat. Am. J. Probab. Math. Stat. {\bf 14} (2017), 631–656. 
%
\bibitem{zuber} J.-B.~Zuber, J. Phys. A: Math. Theor. {\bf 50} (2016) 015203. 
%
\bibitem{spin_flux1} C.~Gattringer, Nucl.Phys. {\bf B850} (2011) 242.  
%
\bibitem{spin_flux2} Y.D.~Mercado, C.~Gattringer, Nucl.Phys. {\bf B862} (2012) 737.  
%
\bibitem{su2pcm_dual} C.~Gattringer, D.~G\"{o}schl, C.~Marchis, Phys.Lett. {\bf B778} (2018) 435.
%
\bibitem{su2_dual} C.~Gattringer, C.~Marchis, Nucl. Phys. {\bf B916} (2017) 627.
%
\bibitem{su3_dual} C.~Marchis, C.~Gattringer,  Phys. Rev. {\bf D97} (2018) 034508. 
%
\bibitem{dual1} O.~Borisenko, V.~Chelnokov, S.~Voloshin, Proc. Lattice 2017, 
EPJ Web of Conferences {\bf 175}, 11021 (2018); arXiv:1712.03064 [hep-lat].  
%
\bibitem{unger_lat18} G.~Gagliardi, W.~Unger, Towards a Dual Representation of Lattice QCD, 
arXiv:1811.02817 [hep-lat].
%
\bibitem{rossi} P.~Rossi, U.~Wolff, Nucl.Phys. {\bf B248} (1984) 105. 
%
\bibitem{strcpl_qcd} G.~Gagliardi, J.~Kim and W.~Unger, Proc. Lattice 2017, 
EPJ Web of Conferences {\bf 175}, 07047 (2018); arXiv:1710.07564 [hep-lat].
%
\bibitem{vicari} S.~Profumo, E.~Vicari, JHEP 0205:014 (2002).  
%

  	
\end{thebibliography}
\end{document}